\documentclass[aps,prb,preprint,superscriptaddress]{revtex4-1}
\usepackage{graphicx}
\usepackage{dcolumn}
\usepackage{bm}
\usepackage{color}
\usepackage{array}

\newcommand{\nn}{\nonumber}
\newcommand{\bb}[0]{\begin{eqnarray}}
\newcommand{\ee}[0]{\end{eqnarray}}

\newcommand{\ff}{\frac{1}{2}}

\begin{document}


\title{Recent developments in the determination of the amplitude and phase of quantum oscillations for the linear chain of coupled orbits}



\author{Alain Audouard}
\email[E-mail address: ]{alain.audouard@lncmi.cnrs.fr}
\affiliation{Laboratoire National des Champs Magn\'{e}tiques
Intenses (UPR 3228 CNRS, INSA, UJF, UPS) 143 avenue de Rangueil,
F-31400 Toulouse, France.}

\author{Jean-Yves~Fortin}
\email[E-mail address: ]{fortin@ijl.nancy-universite.fr}
 \affiliation{Institut Jean Lamour, D\'epartement de Physique de la
Mati\`ere et des Mat\'eriaux,
CNRS-UMR 7198, Vandoeuvre-les-Nancy, F-54506, France.
}%


\date{\today}

\begin{abstract}
De Haas-van Alphen oscillations are studied for Fermi surfaces (FS) illustrating the
model proposed by Pippard in the early sixties, namely the
linear chain of orbits coupled by magnetic breakdown. This FS topology is
relevant for many multiband quasi-two dimensional (q-2D) organic metals such as
$\kappa$-(BEDT-TTF)$_2$Cu(NCS)$_2$ and
$\theta$-(BEDT-TTF)$_4$CoBr$_4$(C$_6$H$_4$Cl$_2$) which are considered in
detail. Whereas the Lifshits-Kosevich model only involves a first order
development of field- and temperature-dependent damping factors, second order
terms may have significant contribution on the Fourier components amplitude for such q-2D systems at
high magnetic field and low temperature. The strength of these second order
terms depends on the relative value of the involved damping factors, which are
in turns strongly dependent on parameters such as the magnetic breakdown field,
effective masses and, most of all, effective Land\'{e} factors. In addition, the influence of
field-dependent Onsager phase factors on the oscillation
spectra is considered.
\end{abstract}


\maketitle

\section{\label{sec:Intro}Introduction}

While the Lifshits-Kosevich (LK) model \cite{LK,Sh84} nicely accounts for de Haas-Van Alphen oscillations spectra relevant to three-dimensional Fermi surfaces (FS), strong deviations are observed for multiband two-dimensional metals, in particular at high magnetic field and low temperature. This is the case, among others, of the starring  $\kappa$-(BEDT-TTF)$_2$Cu(NCS)$_2$ and the recently studied $\theta$-(BEDT-TTF)$_4$CoBr$_4$(C$_6$H$_4$Cl$_2$) charge transfer salts (where BEDT-TTF stands for the bis-ethylenedithio-tetrathiafulvalene molecule). The FS of these organic metals \cite{Ur88,Ju89,Au12,Au13} is an illustration of the textbook model proposed by Pippard more than fifty years ago to compute the Landau band structure induced by magnetic breakdown (MB) in multiband metals \cite{Pi62} (see Fig.~\ref{fig:FS_orbits}). Many experimental studies have demonstrated that such FS topology gives rise to dHvA oscillations spectra composed of linear combinations of the frequencies linked to the basic orbit $\alpha$ and the MB orbit $\beta$ \cite{Me95,Ha96,Uj97,St99,Au12,Au13}. These frequencies correspond not only to actual semiclassical MB orbits, few examples of which are displayed in Fig.~\ref{fig:FS_orbits}, or harmonics, but also to 'forbidden frequencies' such as $\beta-\alpha$ that are not predicted by the semiclassical model of Falicov-Stachowiak \cite{Fa66,Sh84}. Besides, even in the case where these Fourier components correspond to MB orbits, their field and temperature dependence may be at odds with this model. On the other hand, the Onsager phase factor of the oscillations had not been considered until recently though, according to the pioneering works of Slutskin and Kadigrobov
\cite{Sl67} and Kochkin \cite{Ko68}, an additional field-dependent Onsager phase should be
introduced at each Bragg reflection (see Fig.~\ref{fig:FS_orbits}). It is worthwhile to notice that the same result was derived almost ten years later \cite{Hu76} in order to account for the discrepancy
between calculations, which are valid for the low field range, and the
experimental data for the lens orbit of Cd \cite{Co71} which, as it is the case of all the orbits involving $\alpha$ in Fig.~\ref{fig:FS_orbits}, undergoes such Bragg reflections.

Recently, analytic tools have been provided to account for both field and
temperature dependence of the Fourier amplitudes and the Onsager phases relevant to the various frequencies observed \cite{Au12,Au13}. These calculations are first summarized in Section~\ref{sec:model}. Influence of the various physical parameters involved in the oscillations spectra (effective masses, Dingle temperatures, MB field and Land\'{e} factors) on the deviations from the semiclassical model of Falicov-Stachowiak is considered in Section ~\ref{sec:consequences}. To that purpose, the two organic charge transfer salts $\kappa$-(BEDT-TTF)$_2$Cu(NCS)$_2$ and $\theta$-(BEDT-TTF)$_4$CoBr$_4$(C$_6$H$_4$Cl$_2$) are considered.

\section{\label{sec:model}Model}

In this section, we first recall the model accounting for the field and temperature dependence of the amplitude of the various Fourier components entering the oscillation spectra \cite{Au12,Au13}. In the second step, the field-dependent Onsager phase\cite{Au13} is considered.

\subsection{Fourier amplitude}

\begin{figure}
\centering
\includegraphics[width=0.8\columnwidth,clip,angle=0]{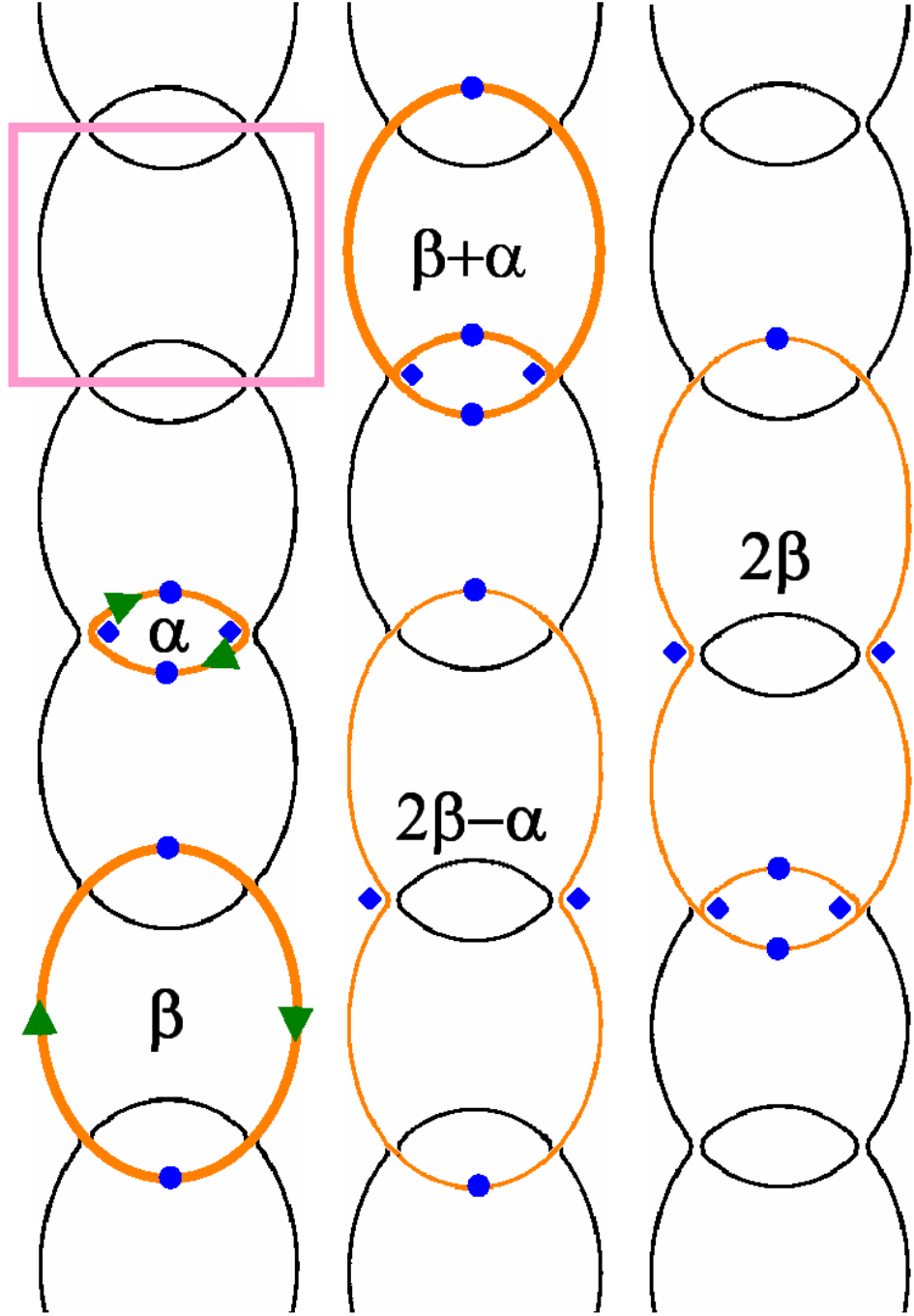}  
\caption{\label{fig:FS_orbits} (Color online) Model Fermi surface relevant to the Pippard's model in the extended zone scheme.
Pink solid lines depict the first Brillouin zone. Orange lines display the
semi-classical orbits considered for the data analysis and arrows indicate the
quasi-particles path on the principal orbits $\alpha$ and $\beta$.
Blue circles mark the turning points in the direction parallel to the chains. Blue diamonds indicate the Bragg reflection points.}
\end{figure}

\begin{figure}
\centering
\includegraphics[width=0.8\columnwidth,clip,angle=0]{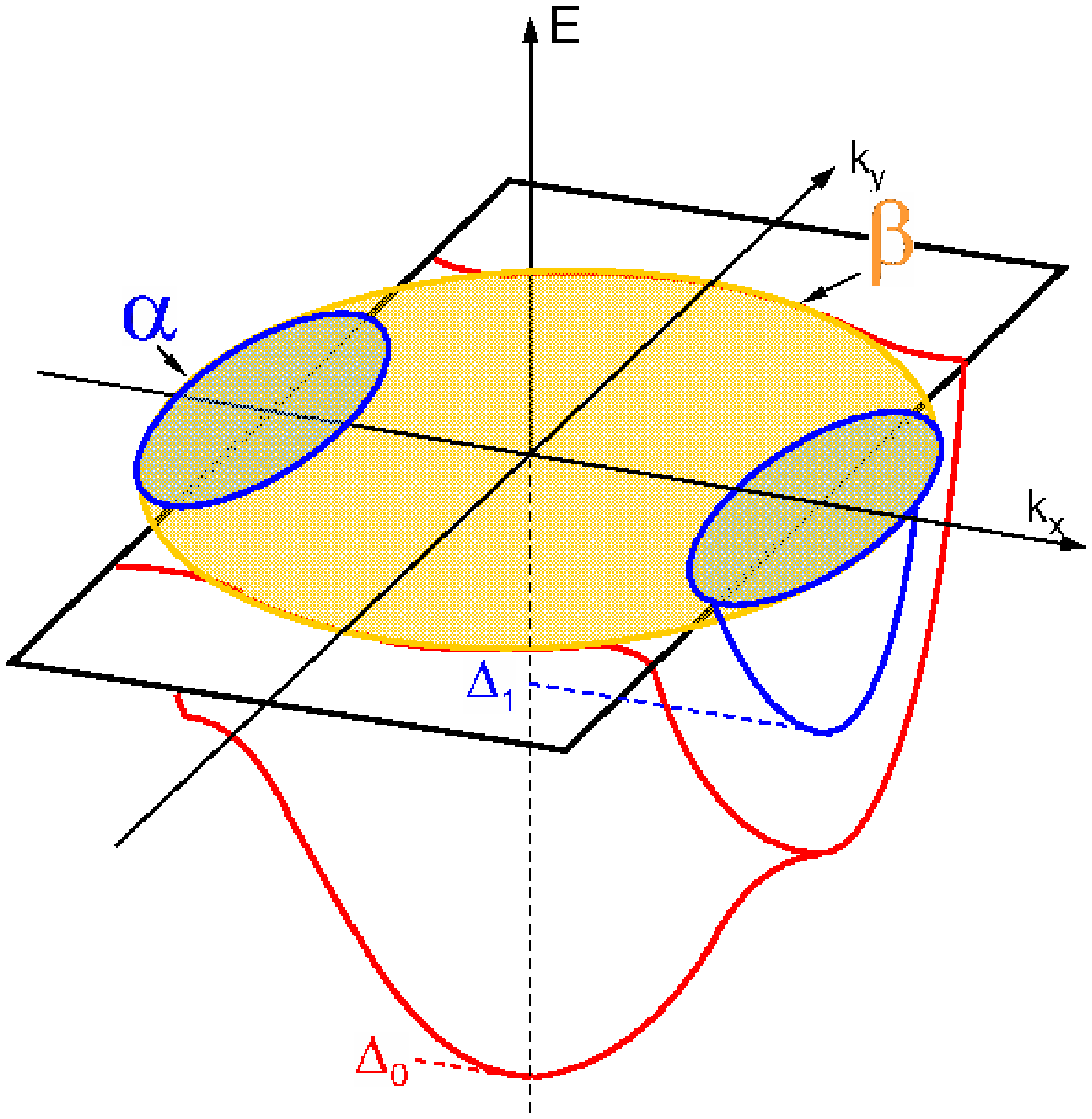}  
\caption{\label{fig:FS_3D} (Color online) Band dispersion scheme relevant to the Pippard's model. Parallelogram in black solid lines depicts the first Brillouin zone. The $\alpha$ orbit in built on band \#1 with bottom energy $\Delta_1$. The magnetic breakdown orbit $\beta$ is built on both band \#0 with bottom energy $\Delta_0$ and band \#1.}
\end{figure}

As displayed in Fig.~\ref{fig:FS_orbits}, the FS is composed of the $\alpha$ quasi-two-dimensional closed tube and a pair of quasi-one-dimensional sheets separated from the $\alpha$ orbit by a gap liable to be overcome by MB.  Numerous semi-classical MB orbits can be defined ($\eta$ = $\alpha$, $\beta$, $\alpha+\beta$, 2$\beta-\alpha$, 2$\beta$, $etc.$), the area of which are linear combinations of those relevant to the  $\alpha$ and  $\beta$ orbits. The area of the latter is equal to that of the first Brillouin zone. Incidentally, it can be remarked that 2$\beta$ corresponds to both the classical orbit displayed in Fig.~\ref{fig:FS_orbits} and the 2$^{nd}$ harmonic of $\beta$.

To account for this FS, a two-band system with band extrema $\Delta_{0(1)}$ and
effective masses $m_{0(1)}$ (in units of the electron mass $m_e$) is considered
\cite{Au12} as reported in Fig.~\ref{fig:FS_3D}. The
band \#0 gives rise to the quasi-one-dimensional part of the FS of
Fig.~\ref{fig:FS_orbits} whereas the $\alpha$ orbit is built on the band \#1. Assuming parabolic dispersion, the relevant frequency  is $F_{\alpha}=m_{1}(\mu-\Delta_{1})$~\cite{fn2}. The $\beta$ orbit, generated by four tunnelings at the junction points, is built on both bands \#0 and \#1 and, still for a parabolic band, has a frequency corresponding to the first Brillouin zone area, $F_{\beta}=m_{\alpha}(\mu-\Delta_{\alpha})+m_{0}(\mu-\Delta_{0})=m_{\beta} (\mu-\Delta_{\beta}).$ In this case, $m_0+m_{\alpha}$ is identified to the mass $m_{\beta}$ of the orbit $\beta$.

To compute the oscillating part of the magnetization at fixed number $N$ of quasi-particles, we need to consider the oscillatory part of the free energy, defined by

\begin{eqnarray}
F(T,N,B)= \Omega(T,\mu,B)+N\mu
\end{eqnarray}

For a constant $N$, the oscillatory part of the grand potential $\Omega$ for a sample slab with area $\cal A$ can be written

\begin{eqnarray}
\nonumber
\phi_0\frac{u_0}{k_B}\frac{\Omega(T,\mu,B)}{\cal A}&=&
-\frac{m_0}{2}(\mu-\Delta_0)^2-\frac{m_{1}}{2}(\mu-\Delta_{1})^2
\\
\label{eq:omega}
&+&\frac{B^2}{2}\sum_{p\ge 1}\sum_{\eta}\frac{C_{\eta}}{\pi^2p^2m_{\eta}}
R_{\eta,p}(B,T)\cos(2\pi pF_{\eta}/B+p\varphi_{\eta}).
\end{eqnarray}

Damping factors can be expressed as $R_{\eta,p}(B,T)$ = $R^T_{\eta,p}(B,T) R^{D}_{\eta,p}(B) R^{MB}_{\eta,p}(B) R^{s}_{\eta,p}$
where:

\begin{eqnarray}
R^{T}_{\eta,p}= pX_{\eta} \sinh^{-1}(pX_{\eta}),\\
\label{Eq:RD}
R^{D}_{\eta,p} = \exp(-pu_0m_{\eta}T_DB^{-1}),\\
\label{Eq:RMB}
R^{MB}_{\eta,p} = (ip_0)^{n^t_{\eta}}(q_0)^{n^r_{\eta}},\\
\label{Eq:Rs}
R^{s}_{\eta,p} = \cos(\pi g^*_{\eta}m_{\eta}/2).
\end{eqnarray}

The field-and temperature-dependent variable ($X_{\eta}$) and the constant ($u_0$) are expressed
as $X_{\eta}$ = $u_0 m_{\eta} T/B$ and $u_0$ = 2$\pi^2 k_B m_e(e\hbar)^{-1}$ = 14.694 T/K. The tunneling ($p_0$) and reflection ($q_0$) probabilities are given by $p_0$ = $e^{-B_0/2B}$
and $p_0^2$ + $q_0^2$ = 1 \cite{Sh84}. $\phi_0=h/e$ is the magnetic flux quantum, $T_D$ is the Dingle temperature
defined by $T_{D}$ = $\hbar(2\pi k_B\tau)^{-1}$, where $\tau^{-1}$ is the
scattering rate, $B_0$ is the MB field, $m_{\eta}$ and $g^*_{\eta}$ are the
effective masses and effective Land\'{e} factor, respectively \cite{fn1}.

Frequencies $F_{\eta}$ can be written as
$F_{\eta}=m_{\eta}(\mu-\Delta_{\eta})$ and are dependent on the
chemical potential $\mu$ since they are proportional to the area enclosed by
the orbits. Coefficients $C_{\eta}$ are the
symmetry factors of orbits $\eta$. Namely,
$C_{\alpha}=C_{\beta}=C_{2\beta-\alpha}=1$ and $C_{\alpha+\beta}=C_{2\beta}=2$.
Integers $n^t_{\eta}$ and $n^r_{\eta}$ are the number of MB-induced tunnelings and reflections, respectively.
$\varphi_{\eta}$ is the Onsager phase factor of the orbit $\eta$, defined by
the number of turning points, $i.e.$ $\pi/2$ times the number of extrema of the
orbit along one direction (see Fig.~\ref{fig:FS_orbits}). $N$ is given by
$d\Omega/d\mu=-N$, and the chemical potential
satisfies the following implicit equation:

\begin{eqnarray}
\label{mu_0}
\nonumber
\mu=\mu_0-\frac{B}{m_{\beta}}\sum_{p\ge 1}\sum_{\eta}\frac{1}{\pi
p}C_{\eta}
R_{\eta,p}(T)\sin(2\pi p\frac{F_{\eta}}{B}+p\varphi_{\eta}),
\end{eqnarray}

which can be rewritten as

\begin{eqnarray}
\label{mu}
\mu = \mu_0-\frac{B}{m_{\beta}}\sum_{\eta}M_{\eta}(B).
\end{eqnarray}

where $\mu_0$ is the zero-field Fermi energy. For a compensated system, in
which case $N=0$, it is equal to
$\mu_0=(m_0\Delta_0+m_{\alpha}\Delta_{\alpha})/(m_0+m_{\alpha})$. The
oscillatory part of the magnetization is defined as

\begin{eqnarray}
\label{Eq:m_osc}
m_{osc}[T]=-\frac{\phi_0 u_0}{{\cal A}k_B}\frac{\partial F(T,N,B)}{\partial B}.
\end{eqnarray}

Solving Eq.
\ref{Eq:m_osc} at the second order in $R_{\eta,p}(B,T)$ (the first order
part corresponding to the LK semi-classical result) yields, after some algebra, to an expansion in power terms of the amplitudes

\begin{eqnarray}\nonumber
m_{osc}&=& -\sum_{\eta}\sum_{p\ge 1}\frac{F_{\eta}C_{\eta}}{\pi p
m_{\eta}}
R_{\eta,p}(B,T)\sin\left ( 2\pi p \frac{F_{\eta}}{B}+p\varphi_{\eta} \right )
\\ \nonumber
&+&\sum_{\eta,\eta'}\sum_{p,p'\ge 1}\frac{F_{\eta}C_{\eta}C_{\eta'}}{\pi p' m_{\beta}}
R_{\eta,p}(B,T)R_{\eta',p'}(B,T)
\left [
\sin\left ( 2\pi \frac{pF_{\eta}+p'F_{\eta'}}{B}+p\varphi_{\eta}+p'\varphi_{\eta'} \right )
\right .
\\ \label{Eq:m_osc2}
&-&\left .\sin\left ( 2\pi \frac{pF_{\eta}-p'F_{\eta'}}{B}+p\varphi_{\eta}-p'\varphi_{\eta'} \right )
\right ]+\cdots
\end{eqnarray}

where the next terms are third order. From this step onwards, frequencies $F_{\eta}$ are evaluated
at $\mu=\mu_0$: $F_{\eta}=m_{\eta}(\mu_0-\Delta_{\eta})$. According to the above
expression, magnetization spectrum can be expressed in terms of both
classical and non-classical frequencies, still noted as $F_{\eta}$ in the
following, and can be expanded as:

\begin{eqnarray}
\label{Eq:m_osc3}
m_{osc}=\sum_{\eta,p\ge 1}A_{p\eta}\sin\left (2\pi p \frac{F_{\eta}}{B}+p\phi_{\eta} \right ).
\end{eqnarray}

It is important to stress that the amplitude $A_{p\eta}$ involves not only the contribution of
the $p^{th}$ harmonics of the $\eta$ classical orbit, given by the LK formalism
($A_{p\eta}\propto R_{\eta,p}) $ but also higher order corrections,
calculated here at the second order in damping factors. The expressions of the
dominant Fourier components, considered for the data analysis, are:

\begin{eqnarray}
\label{Eq:alpha}
A_{\alpha}&=&-\frac{F_{\alpha}}{\pi m_{\alpha}}R_{\alpha,1}-
\frac{F_{\alpha}}{\pi m_{\beta}}
\left [
\ff R_{\alpha,1}R_{\alpha,2}
+\frac{1}{6}R_{\alpha,2}R_{\alpha,3}+2R_{\beta,1}R_{\alpha+\beta,1}
+\ff R_{\beta,2}R_{2\beta-\alpha,1}
\right ]
\\ 
\label{Eq:2alpha}
A_{2\alpha}&=&-\frac{F_{\alpha}}{2\pi m_{\alpha}}R_{\alpha,2}+
\frac{F_{\alpha}}{\pi m_{\beta}}\left [
R_{\alpha,1}^2-\frac{2}{3}R_{\alpha,1}R_{\alpha,3}-R_{\alpha,2}
R_{\alpha+\beta,2} \right ]
\\ 
\label{Eq:beta}
A_{\beta}&=&-\frac{F_{\beta}}{\pi m_{\beta}}R_{\beta,1}-
\frac{F_{\beta}}{\pi m_{\beta}}
\left [
\ff R_{\beta,1}R_{\beta,2}
+\frac{1}{6}R_{\beta,2}R_{\beta,3}+2R_{\alpha,1}R_{\alpha+\beta,1}
+2R_{\beta,1}R_{2\beta,1}
\right ]
\\ \nn
\label{Eq:2beta}
A_{2\beta}&=&-\frac{F_{\beta}}{2\pi m_{\beta}}\left
[R_{\beta,2}+2R_{2\beta,1}\right ]+
\frac{F_{\beta}}{\pi m_{\beta}}
\left [
R_{\beta,1}^2-\frac{2}{3}R_{\beta,1}R_{\beta,3}-\frac{1}{4}R_{\beta,2}R_{\beta,4
}-R_{\alpha,2}R_{\alpha+\beta,2}\right .
\\ 
&+&\left .2R_{\alpha,1}R_{2\beta-\alpha,1}-R_{\beta,2}R_{
2\beta,2}-R_{\beta,4}R_{2\beta,1}
\right ]
\\ 
\label{Eq:beta-alpha}
A_{\beta-\alpha}&=&-\frac{F_{\beta-\alpha}}{\pi m_{\beta}}\left [
R_{\alpha,1}R_{\beta,1}+R_{\alpha,2}R_{\alpha+\beta,1}+R_{\beta,2}R_{
\alpha+\beta,1}+R_{\beta,1}R_{2\beta-\alpha,1}\right
]
\\ 
\label{Eq:beta+alpha}
A_{\beta+\alpha}&=&-\frac{2F_{\beta+\alpha}}{\pi m_{\beta+\alpha}}
R_{\beta+\alpha,1}+\frac{F_{\beta+\alpha}}{\pi
m_{\beta}}\left [R_{\alpha,1}R_{\beta,1}-2R_{\alpha+\beta,2}R_{\alpha+\beta,1}
-\frac { 1}{3}R_ { \beta,3}R_{2\beta-\alpha,1} \right
]
\\
\label{Eq:2beta-alpha}
A_{2\beta-\alpha}&=&-\frac{F_{2\beta-\alpha}}{\pi m_{2\beta-\alpha}}
R_{2\beta-\alpha,1}-
\frac{F_{2\beta-\alpha}}{\pi m_{\beta}}
\left [\ff R_{\alpha,1}R_{\beta,2}+
\frac{1}{3}
R_{\alpha,3}R_{\alpha+\beta,2}\right ]
\\ \nn
\label{Eq:2beta-2alpha}
A_{2\beta-2\alpha}&=&-\frac{F_{2\beta-2\alpha}}{\pi m_{\beta}}
\left [2R_{\alpha,2}R_{2\beta,1}+2R_{2\beta-\alpha,2}R_{2\beta,1}
+2R_{\alpha,1}R_{2\beta-\alpha,1}+\frac{1}{2}R_{\alpha,4}R_{\alpha+\beta,2}
\right .
\\
&+&\left .
\frac{1}{2}R_{\alpha,2}R_{\beta,2}+\frac{1}{2}R_{\beta,2}R_{2\beta-\alpha,2}
\right ].
\end{eqnarray}

The leading term of Eqs.~\ref{Eq:alpha}, \ref{Eq:2alpha}, \ref{Eq:beta}, \ref{Eq:beta+alpha} and \ref{Eq:2beta-alpha} corresponds to the LK formalism. In the specific case of Eq.~\ref{Eq:2beta}, it involves the contributions of both the classical
orbit $2\beta$ displayed in Fig.~\ref{fig:FS_orbits} and the second harmonics of $\beta$ which are accounted for by
the damping factors $R_{2\beta,1}$ and $R_{\beta,2}$, respectively. In contrast, there is no first order term
entering Eqs.~\ref{Eq:beta-alpha} and \ref{Eq:2beta-2alpha} relevant to $\beta-\alpha$ and its second harmonics,
respectively, since these Fourier components correspond to 'forbidden frequencies'. Since, generally speaking, these equations involve algebraic sums and products of damping factors, care must be taken in their sign. Namely, the sign of the spin damping factor, which is the only one liable to be negative according to Eq.~\ref{Eq:Rs}, must be taken into account.

\subsection{Onsager phase factor}

\begin{figure}
\centering
\includegraphics[width=0.8\columnwidth,clip,angle=0]{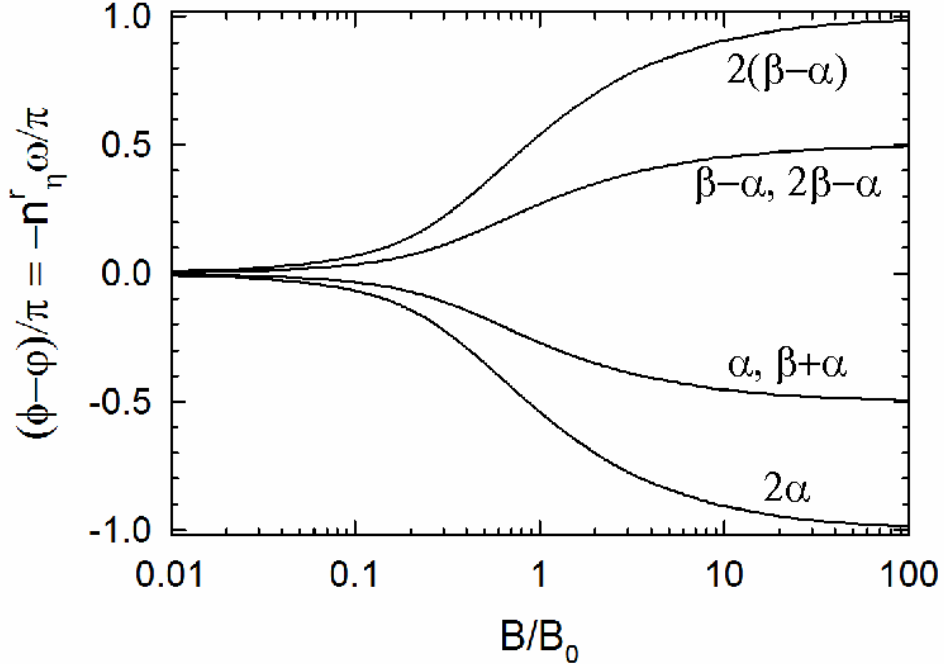}
\caption{\label{fig:omega} Field-dependent part of the Onsager phase given by Eqs.~\ref{Eq:matrix} to~\ref{Eq:phi} for the orbits $\eta$ considered in Eqs.~\ref{Eq:alpha} to~\ref{Eq:2beta-2alpha}. For small fields $B$ compared to the magnetic breakdown field
$B_0$, the field-dependent part vanishes while it goes to $n^r_{\eta}\pi/4$ at large fields, where $n^r_{\eta}$ is the number of Bragg reflections encountered by the quasi-particles during its path on the orbit $\eta$.}
\end{figure}

Turn on now to the determination of the Onsager phase factors $\phi_{\eta}$
entering Eq.~\ref{Eq:m_osc3}. Within the semiclassical theory, a phase factor
$\pi/2$ is introduced at each turning point (see blue circles in
Fig.~\ref{fig:FS_orbits}) leading to the phase factor  $\varphi_{\eta}$
appearing in Eq.~\ref{eq:omega}. More specifically,
$\varphi_{\eta}$ = $2\pi$ for $\eta$ = $2\alpha$,
$2\beta$, $\beta+\alpha$, $2(\beta+\alpha)$ and $\pi$ for $\eta$ = $\alpha$,
$\beta$, $2\beta-\alpha$. In addition to this phase factor, an additional field-dependent
phase $\omega$ is added to $\varphi_{\eta}$ each time a quasiparticle is
reflected at a MB
junction  (see blue diamonds in Fig.~\ref{fig:FS_orbits}). Indeed, according to Refs. \cite{Sl67,Ko68,Hu76}, the matrix for the
incoming and outgoing wave-function amplitudes at each junction point is given by

\begin{equation}
\label{Eq:matrix}
M =
\left (
\begin{array}{cc}
q_0{\rm e}^{-i\omega} & ip_0 \\
ip_0 & q_0{\rm e}^{i\omega} \\
\end{array}
\right )
\end{equation}

where

\begin{equation}
\omega(B)=-\frac{\pi}{4}+x\ln(x)-x-\arg \Gamma(ix),\;\;
x=\frac{B_0}{2\pi B}.
\end{equation}

After a reflection, the quasi-particle amplitude takes a factor
$q_0\exp(-i\omega)$ and $q_0\exp(i\omega)$ for quasi-particle path orientation clockwise and counter-clockwise, respectively.
Even though $\omega$ goes to zero at low field, it takes noticeable
values as the field is larger than $B_0$, going to $\pi$/4 at large field. According to Eq.~\ref{Eq:matrix}, the Onsager phase factor is given by

\begin{equation}
\label{Eq:phi}
\phi_{\eta} = \varphi_{\eta} - n^r_{\eta}\omega(B)
\end{equation}

where $n^r_{\eta}$ = 2 for $\eta$ = $\alpha$ and $\beta+\alpha$,  $n^r_{\eta}$ = -2  for  $\eta$ = $\beta-\alpha$ and $2\beta-\alpha$, $n^r_{\eta}$ = 4  for  $\eta$ =  $2\alpha$, $n^r_{\eta}$ = -4  for  $\eta$ =  $2(\beta-\alpha)$ and  $n^r_{\eta}$ = 0 for $\eta$ = $\beta$ and $2\beta$ since these latter components only involve tunnelings. In that respect, it can be remarked that even though the Fourier component with frequency $F_{2\beta}$ arises from both the second harmonics of $\beta$ and the 2$\beta$ orbit displayed in Fig.~\ref{fig:FS_orbits}, these two contributions have the same Onsager phase. Besides, for a given $\eta$ Fourier component, all the involved second order terms (see Eqs.~\ref{Eq:alpha} to \ref{Eq:2beta-2alpha}) can be viewed as arising from algebraic combinations of classical orbits yielding the same Onsager phase. As a consequence, the index $n^r_{\eta}$ can be negative, due
to algebraic combinations of the individual phases present in the sine function
of Eq.~\ref{Eq:m_osc2}. The field dependence of $\phi_{\eta}$ reported in Fig.~\ref{fig:omega} demonstrates that, excepted for  $\beta$ and $2\beta$, significant Onsager phase shifts should be observed at high $B/B_0$ ratio.

\section{\label{sec:consequences}Consequences for oscillations spectra}

\begin{figure}
\includegraphics[width=0.8\columnwidth,clip,angle=0]{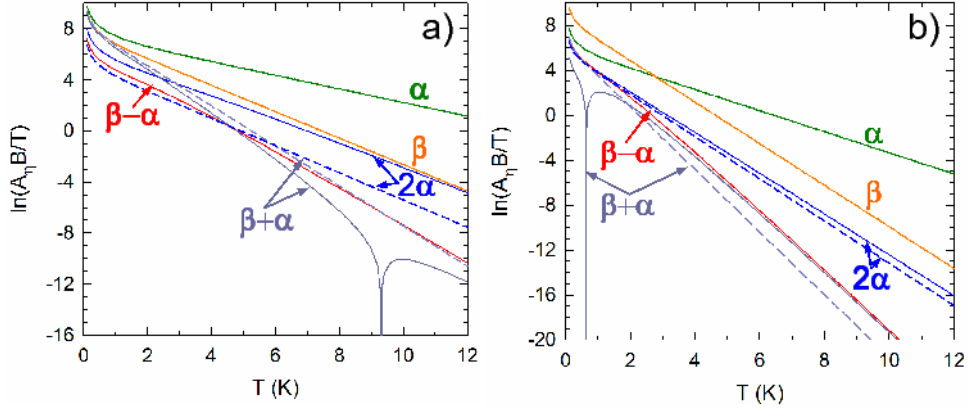}
\caption{\label{Fig:massplot} Temperature dependence, at 50 T, of the Fourier amplitude linked to few of the components relevant to Eqs.~\ref{Eq:alpha} to \ref{Eq:2beta-2alpha} (solid lines) and corresponding predictions of the Lifshits-Kosevich formalism (dashed lines). (a) Data for $\theta$-(BEDT-TTF)$_4$CoBr$_4$(C$_6$H$_4$Cl$_2$). The frequencies are $F_{\alpha}$ = 944 T and  $m_{\beta}$ = 4600 T. Effective masses are $m_{\alpha}$ = 1.81 and  $m_{\beta}$ = 3.52. The effective Land\'{e} factors are $g^*_\alpha$ = $g^*_\beta$ = 1.9. The Dingle temperature and MB field are $T_{D\alpha}$ = $T_{D\beta}$ = 0.79 K and $B_0$ = 35 T. (b) Data for  $\kappa$-(BEDT-TTF)$_2$Cu(NCS)$_2$. The frequencies are $F_{\alpha}$ = 600 T and  $F_{\beta}$ = 3900 T. Effective masses are $m_{\alpha}$ = 3.2 and  $m_{\beta}$ = 6.3. The effective Land\'{e} factors are $g^*_\alpha$ = 1.5 and $g^*_\beta$ = 1.6. The Dingle temperature and MB field are $T_{D\alpha}$ = $T_{D\beta}$ = 0.5 K and $B_0$ = 25 T.}
\end{figure}

\begin{figure}
\includegraphics[width=0.8\columnwidth,clip,angle=0]{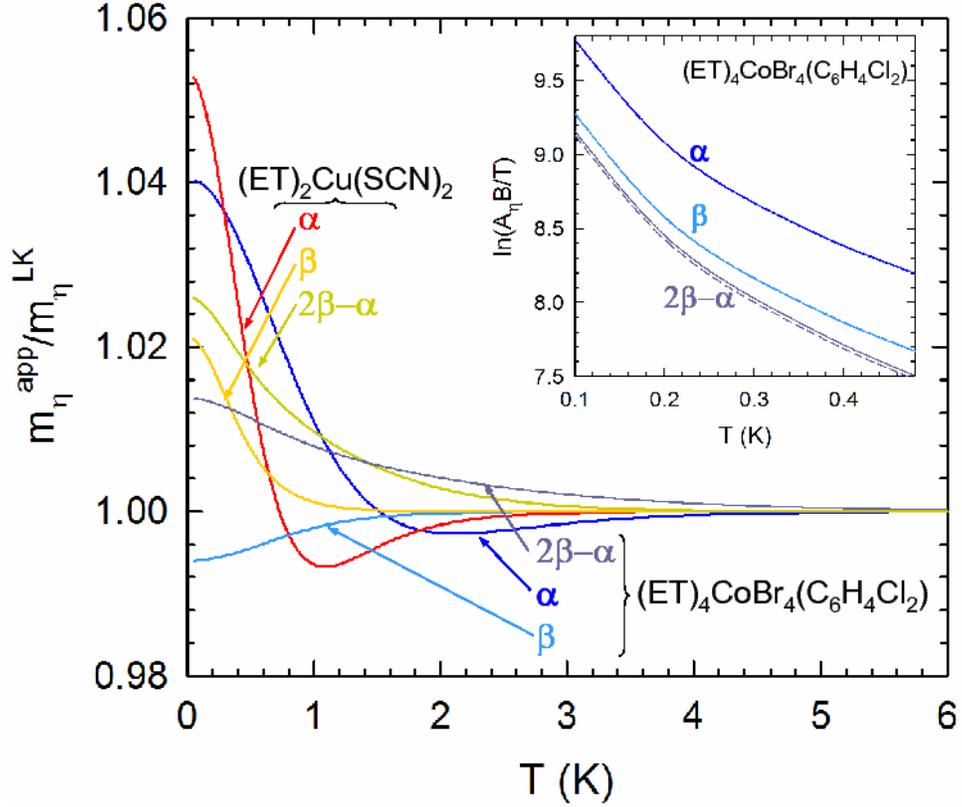}
\caption{\label{Fig:mapp_semiclassiques} Temperature dependence of the apparent effective masses calculated from Eq.~\ref{Eq:mapp} for the $\alpha$, $\beta$ and $2\beta-\alpha$ semiclassical orbits of  $\kappa$-(BEDT-TTF)$_2$Cu(NCS)$_2$ and $\theta$-(BEDT-TTF)$_4$CoBr$_4$(C$_6$H$_4$Cl$_2$). Data are deduced from the mass plots reported in the insert and normalized to the Lifshits-Kosevich  predictions.}
\end{figure}

\begin{figure}
\includegraphics[width=0.8\columnwidth,clip,angle=0]{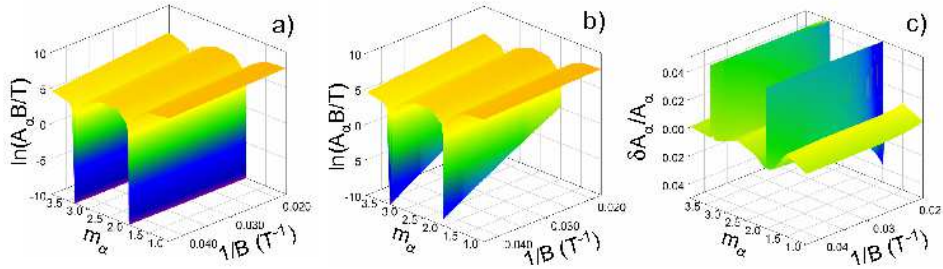}
\caption{\label{Fig:alpha} Field and temperature dependence of the Fourier amplitude $A_{\alpha}$ in the framework of (a) the Lifshits-Kosevich (LK) model and (b) Eq.\ref{Eq:alpha}. The discrepancy between LK and Eq.\ref{Eq:alpha} is given in (c).}
\end{figure}

\begin{figure}
\includegraphics[width=0.8\columnwidth,clip,angle=0]{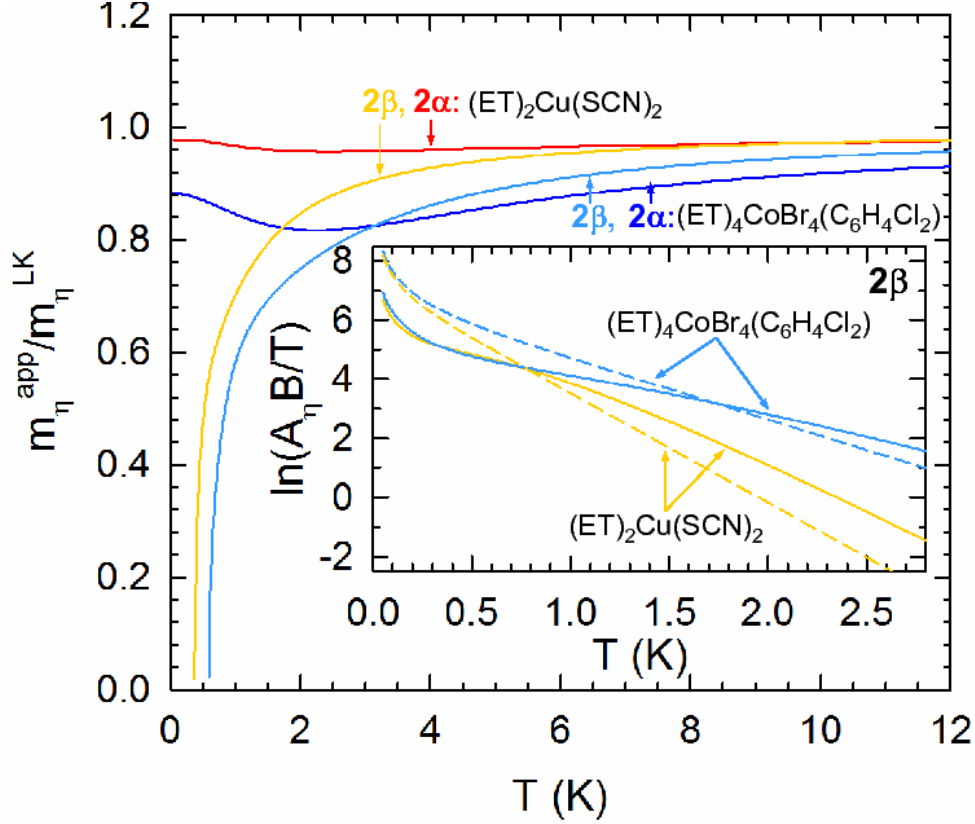}
\caption{\label{Fig:mapp_harmoniques} Same as Fig.~\ref{Fig:alpha} for the harmonics  $2\alpha$ and $2\beta$. The insert compares the predictions of the LK model (dotted lines) and Eq.~\ref{Eq:2beta} (solid lines) for the mass plots of $2\beta$.}
\end{figure}

\begin{figure}
\includegraphics[width=0.8\columnwidth,clip,angle=0]{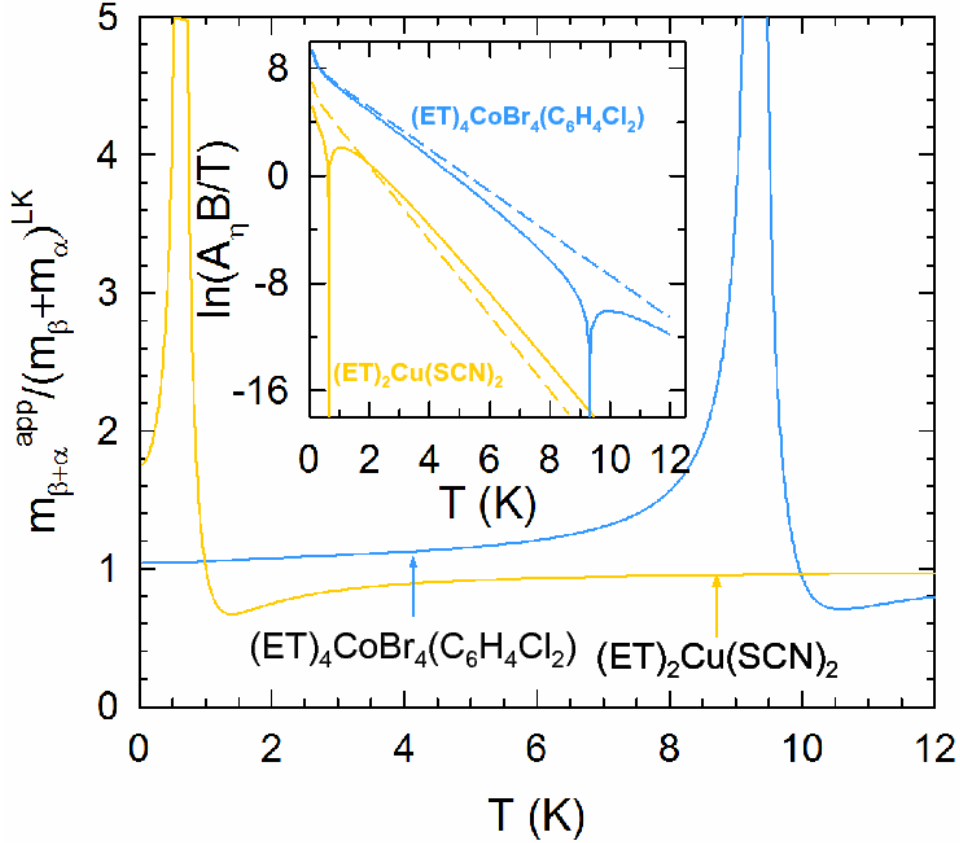}
\caption{\label{Fig:mapp_beta_plus_alpha} Same as Fig.~\ref{Fig:mapp_harmoniques} for the magnetic breakdown orbit $\beta+\alpha$.}
\end{figure}

\begin{figure}
\includegraphics[width=0.8\columnwidth,clip,angle=0]{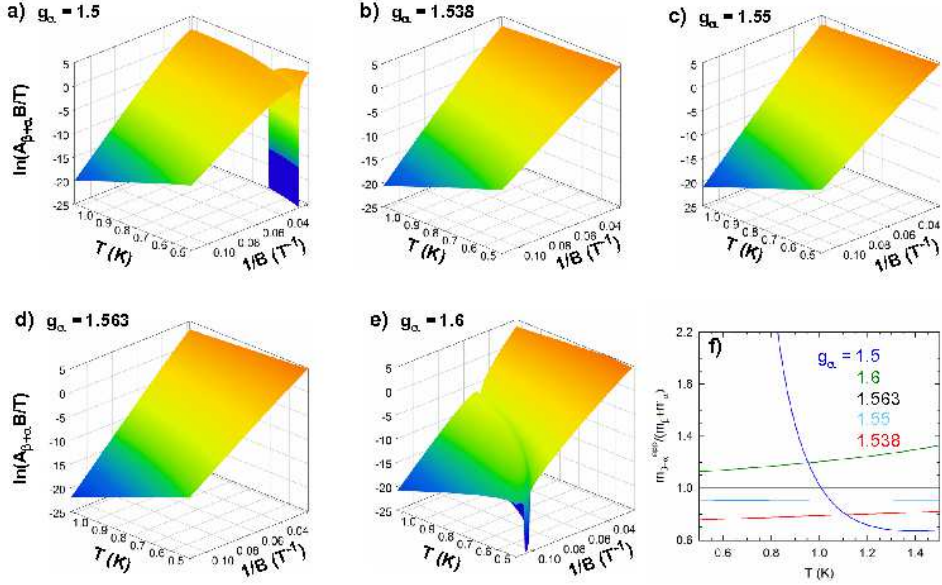}
\caption{\label{Fig:beta+alpha} Field and temperature dependence of the Fourier amplitude linked to the magnetic breakdown (MB) orbit $\beta+\alpha$ predicted by Eq.~\ref{Eq:beta+alpha}. In (a) parameters are the same as in Fig.~\ref{Fig:massplot}. In particular, the effective Land\'{e} factor is $g^*_{\alpha}$ = 1.5. The first and second terms of Eq.~\ref{Eq:beta+alpha} cancel each other at specific couples of temperature and field values yielding a strong dip line. As $g_{\alpha}$ increases, the dip goes towards high fields. (b) For $g^*_{\alpha}$ = 1.538,   $R^s_{\beta+\alpha} =0$ and the amplitude is mainly governed by the second order term involving the product $R_{\alpha}R_{\beta}$. (c) For  $g^*_{\alpha}$ = 1.55, $R_{\beta+\alpha}$ and $R_{\alpha}R_{\beta}$ have an opposite sings. Hence, their difference never cancels. (d) For $g_{\alpha}$ = 1.563, the field dependence is dominated by the first order term since $R^s_{\alpha} = 0$. Therefore, the Lifshits-Kosevich behaviour is recovered. As  $g^*_{\alpha}$ further increases, e.g. for (e) $g^*_{\alpha}$ = 1.6, a behaviour similar to that of (a) is obtained. (f) Apparent effective mass value deduced through Eq.~\ref{Eq:mapp} from the data in (a) to (e) at $B$ = 50 T.}
\end{figure}

\begin{figure}
\includegraphics[width=0.8\columnwidth,clip,angle=0]{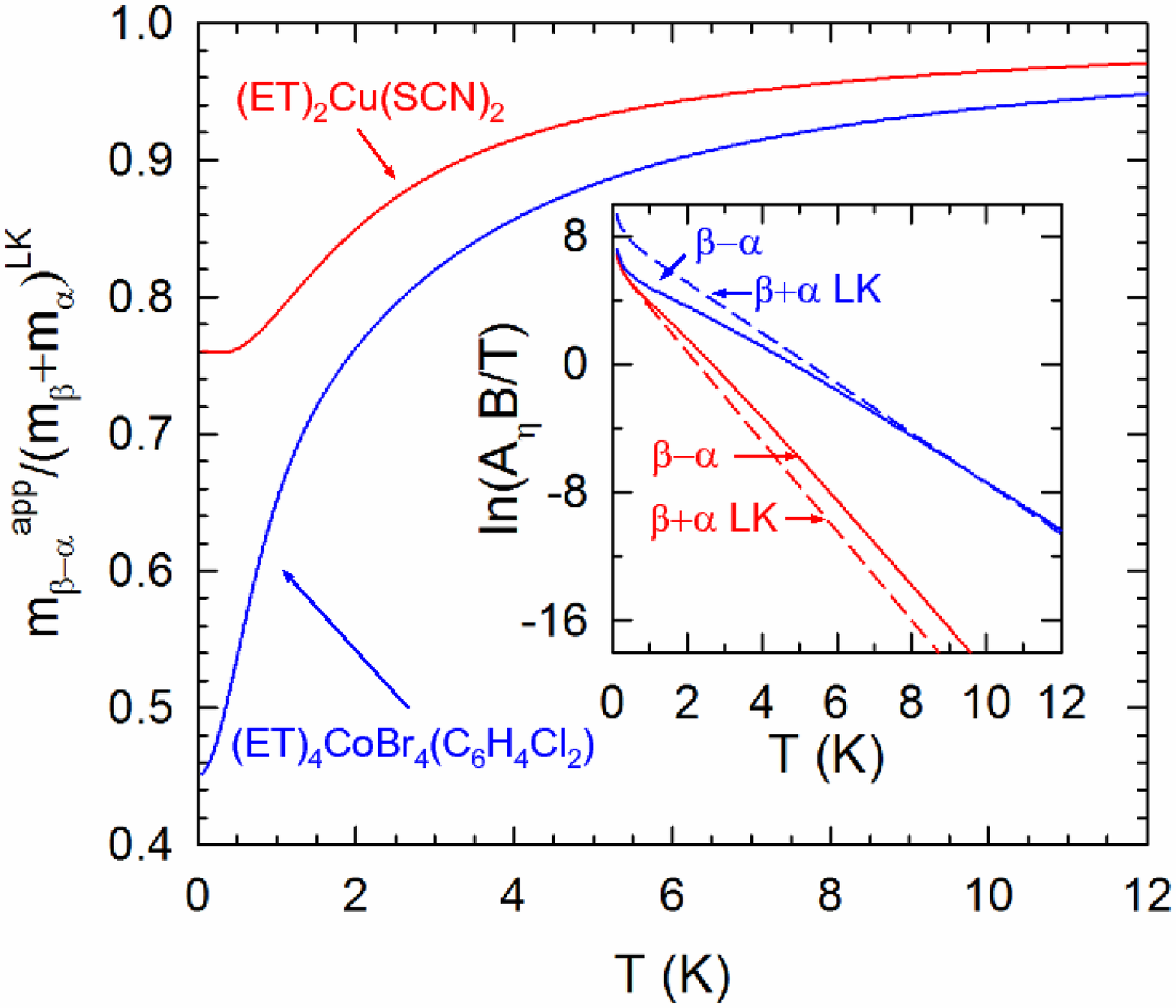}
\caption{\label{Fig:mapp_beta_moins_alpha} Same as Fig.~\ref{Fig:mapp_harmoniques} for the 'forbidden orbit' $\beta-\alpha$. The insert compares the predictions of the LK model for $\beta+\alpha$ (dashed lines) and Eq.~\ref{Eq:beta-alpha} for $\beta-\alpha$ (solid lines). Same slope, hence same effective mass is obtained in both cases at high temperature.}
\end{figure}

Calculations reported in Section~\ref{sec:model} account for dHvA oscillations of $\theta$-(BEDT-TTF)$_4$CoBr$_4$(C$_6$H$_4$Cl$_2$) in the temperature range 1.5 - 4.2 K in magnetic fields of up to 55 T \cite{Au12,Au13}. In particular, the observed deviations from the LK model, including the field and temperature dependence of the 'forbidden orbit' $\beta-\alpha$, and the field-dependent shift of the Onsager phase are reproduced by the model. Nevertheless, as discussed below, the predicted behaviour is strongly dependent on the various parameters involved (effective masses, Land\'{e} factors, MB field, etc.). For this reason, further experiments on other compounds are needed to more extensively check the model. While these experiments are being performed, the influence of the various parameters can be examined. In the following, we consider, as realistic starting points, the parameters relevant to the well known $\kappa$-(BEDT-TTF)$_2$Cu(NCS)$_2$ and the recently studied $\theta$-(BEDT-TTF)$_4$CoBr$_4$(C$_
6$H$_4$Cl$_2$) organic metals.

According to Eqs.~\ref{Eq:alpha} to~\ref{Eq:2beta-2alpha}, the physical parameters involved in the Fourier components' amplitude are the frequencies $F_{\alpha}$ and $F_{\beta}$, effective masses $m_{\alpha}$ and $m_{\beta}$, Dingle temperatures $T_{D\alpha}$ and $T_{D\beta}$, Land\'{e} factors $g^*_{\alpha}$ and $g^*_{\beta}$ and MB field $B_0$. This set of parameters governs the whole field and temperature dependence of the oscillatory spectrum within one constant prefactor ($\tau_0$, see below).

For $\kappa$-(BEDT-TTF)$_2$Cu(NCS)$_2$  \cite{Me95,Ha96,Uj97,St99,Sa90,An91,He91,Ca94,Gv04}, reported frequency values are in the range $F_{\alpha}$ = 597 $\div$ 625 T and $F_{\beta}$ = 3800 $\div$ 3920 T. Effective masses are $m_{\alpha}$ = 3.0 $\div$ 3.5 and $m_{\beta}$ = 5.6 $\div$ 7.1. Very scattered values of the MB field, ranging from $B_0$ = 17 to $B_0$ = 41 T, are deduced from the data. Land\'{e} factors are  $g^*_{\alpha}$ = 1.5 and $g^*_{\beta}$ = 1.6. Dingle temperatures, which are the only sample-dependent parameters are close to 0.5 K within a few tenth of a kelvin. As for $\theta$-(BEDT-TTF)$_4$CoBr$_4$(C$_6$H$_4$Cl$_2$), these parameters are  $F_{\alpha}$= 944 $\pm$ 4 T, $F_{\beta}$ = 4600 $\pm$ 10 T,  $m_{\alpha}$ = 1.81 $\pm$ 0.05, $m_{\beta}$ = 3.52 $\pm$ 0.19, $B_0$ = 35 $\pm$ 5 T, $g^*_{\alpha}$ = $g^*_{\beta}$ = 1.9 $\pm$ 0.2, $T_D$ = 0.79 $\pm$ 0.10 K\cite{Au12,Au13}.

DHvA oscillations are generally deduced from magnetic torque measurements. In
such a case, oscillatory torque amplitudes $A^{\tau}_{\eta}$ are related to dHvA
amplitudes as  $A^{\tau}_{\eta} = \tau_0A_{\eta}B \tan(\theta)$ where $\theta$
is the angle between the field direction and the normal to the conducting plane.
Besides, at high $T/B$ ratio, $\ln(A_{\eta}B/T)$ varies linearly with both the
inverse magnetic field, at a given temperature (Dingle plot), and temperature,
at a given magnetic field (mass plot), in the framework of the LK model. For
these reasons the quantity $\ln(A_{\eta}B/T)$ is considered throughout the
following.

Fig.~\ref{Fig:massplot} compares predictions of the LK formalism and Eqs.~\ref{Eq:alpha} to~\ref{Eq:2beta-2alpha} for mass plots at $B$ = 50 T with parameters relevant to  $\theta$-(BEDT-TTF)$_4$CoBr$_4$(C$_6$H$_4$Cl$_2$) and  $\kappa$-(BEDT-TTF)$_2$Cu(NCS)$_2$. A first finding is that the two models yield close data for the basic orbits $\alpha$ and $\beta$ while discrepancies are observed for the harmonics $2\alpha$. Even stronger deviations are observed for $\beta+\alpha$ , the field dependence of which exhibits a profound dip for both compounds, albeit at a different temperature. Such a behaviour may have significant consequences on the determination of effective masses. For example, according to the data of \cite{Uj97}, $m_{\beta}$ + $m_{\alpha}$ = 9.0 even though $m_{\beta+\alpha}$ = 6.6, only, in violation of the Falicov-Stachowiak model predictions. As pointed out in \cite{Fo09}, the Lifshits-Koscvich formula can be rewritten:

\begin{equation}
\label{Eq:y}
y_{\eta} = \frac{R(B)}{\sinh(m_{\eta}x)}
\end{equation}

where $y_{\eta} = A_{\eta}B/u_0T$, $x = u_0T/B$ and $R(B)$ stands for the temperature-independent contribution of the damping factors, namely $R(B)=-F_{\eta}R^D_{\eta}R^{MB}_{\eta}R^s_{\eta}/\pi$ (see Eqs.~\ref{Eq:RD} to~\ref{Eq:Rs}). At a given magnetic field, an 'apparent' effective mass $m^{app}_{\eta}$ can be extracted from Eq.~\ref{Eq:y} as:

\begin{equation}
\label{Eq:mapp}
m^{app}_{\eta} = \left[-\frac{1}{y} \frac{d^2y}{dx^2}+2 \left(\frac{1}{y} \frac{dy}{dx} \right)^2 \right]^{\frac{1}{2}}
\end{equation}

 At a given field and temperature, the 'local value' of $m^{app}_{\eta}$ can be deduced from mass plots, on the basis of the above equation \cite{fn3}. Of course, in the case where the LK model is actually valid, Eq.~\ref{Eq:mapp} yields $m^{app}_{\eta}$ = $m_{\eta}$. As reported in Figs.~\ref{Fig:massplot} and~\ref{Fig:mapp_semiclassiques}, only a slight discrepancy with the LK model is observed at low temperature for the basic and MB orbits $\alpha$, $\beta$  and  $2\beta-\alpha$. This is due to the very small value of the second order terms compared to the first order term in Eqs.~\ref{Eq:alpha}, \ref{Eq:beta} and \ref{Eq:2beta-alpha}. As a result, the apparent effective masses remain very close to the LK predictions, within few percent, down to very low temperature. 

Nevertheless, the influence of the spin Damping factor on the amplitude must be
taken into account. In the framework of the LK model, a zero amplitude is
obtained for $R^s_{\eta}=0$, $i.e.$ $g^*_\eta m_\eta/(2\cos\theta)=\pi(n+1/2)$
where $n$ is an integer. Even though this feature, known as the spin-zero
phenomenon, is generally experimentally studied through the angle dependence of
the amplitude, such spin zeroes can be obtained by varying the effective mass,
as well. For example, this could be achieved by applying pressure
\cite{Ca94,Au06,Au11,fn4}. Considering the $\alpha$ orbit as an example, strong
dips are actually observed within the LK model at $m_{\alpha}$ values
corresponding to spin zeroes (see Fig.~\ref{Fig:alpha}). Same feature is
predicted by Eq.~\ref{Eq:alpha} since, as already mentioned, the second order
terms of Eq.~\ref{Eq:alpha} are small compared to the first order term. However,
since these high order terms have no reason to cancel at the same spin zero
values as those relevant to $R^s_{\alpha}$, the observed field dependence within the dips, governed by these high order terms, is very strong. Nevertheless, as above mentioned and evidenced in Fig.~\ref{Fig:alpha}(c), only at most few percent of discrepancy is observed far from the zeroes between the LK model and Eq.~\ref{Eq:alpha}. As a consequence, it can be concluded that the LK formalism conveniently accounts for the data relevant to basic orbits, provided their spin damping factors are not too small. This result is important since it indicates that the values reported in the literature for the basic orbits, using the LK formula, are generally valid.

Data for harmonics $2\alpha$ and $2\beta$ reported in Figs.~\ref{Fig:massplot} and~\ref{Fig:mapp_harmoniques} exhibit clear deviations from the LK behaviour, in particular at low temperature, hence strong apparent deviations from the LK effective masses. This behaviour is mainly due to the second order terms $R^2_{\alpha,1}$ and $R^2_{\beta,1}$  (see Eqs.~\ref{Eq:2alpha} and ~\ref{Eq:2beta})  which are of the same order of magnitude as the leading terms $R_{\alpha,2}$ and $R_{\beta,2}$, respectively. However, as pointed out\cite{Au12}, the respective strength of these first and second order terms are strongly dependent on the involved spin damping factors, hence on the effective Land\'{e} factors.

The influence of the effective Land\'{e} factor, or in fact of the product
$g_{\eta}m_{\eta}/\cos(\theta)$, is further evidenced in the case of
$\beta+\alpha$, the field-dependent amplitude of which exhibits a profound dip
(see Fig.~\ref{Fig:mapp_beta_plus_alpha}). At variance with the spin-zero
phenomenon  observed for basic orbits, this dip is due to the cancellation of
the first and second order terms of Eq.~\ref{Eq:beta+alpha}. Indeed, putting
aside the spin damping factors contribution, $R_{\beta+\alpha}$ is close to the
product $R_{\beta}R_{\alpha}$ appearing in Eq.~\ref{Eq:beta+alpha}.
Nevertheless, in line with the data of Fig.~\ref{Fig:beta+alpha}, their respective value
strongly depend on the effective Land\'{e} factors. Otherwise, despite strong
fluctuations around the dips, the LK behaviour, hence the prediction of the
Falicov-Stachowiak model ($m_{\alpha}+m_{\beta} = m_{\beta+\alpha}$) is
recovered at high temperature. It must be pointed out that the occurrence of
such dips has not been reported yet in
experiments. This is not surprising in the case of
$\theta$-(BEDT-TTF)$_4$CoBr$_4$(C$_6$H$_4$Cl$_2$) since the temperature range
explored is below 4.2 K whereas the dip is observed at 9.3 K in
Fig.~\ref{Fig:mapp_beta_plus_alpha}.  According to the data in
Fig.~\ref{Fig:beta+alpha}, the behaviour of $\kappa$-(BEDT-TTF)$_2$Cu(NCS)$_2$
is strongly dependent on the value of $g_{\alpha}$ (the $g_{\alpha}$ range
explored in Fig.~\ref{Fig:mapp_beta_plus_alpha} remains within the reported
experimental uncertainty). It must be pointed out that changing the value of
$g_{\alpha}$ is equivalent to change the value of
$g_{\beta+\alpha}m_{\beta+\alpha}/\cos(\theta)$, hence the value of the tilt
angle $\theta$. Such angle differences could explain the discrepancy between the
data of Refs.~\onlinecite{Me95} (for which $m_{\beta+\alpha} =
m_{\alpha}+m_{\beta}$, in agreement with the data in
Fig.~\ref{Fig:beta+alpha}(d)) and~\onlinecite{Uj97} (where
$m_{\beta+\alpha}/(m_{\alpha}+m_{\beta}) = 0.73 $, which is more in line with
the data of Fig.~\ref{Fig:beta+alpha}(e)).

To conclude with mass plots let us consider the 'forbidden orbit'  $\beta-\alpha$, the data of which are reported in Fig.~\ref{Fig:mapp_beta_moins_alpha}. It can be remarked first that the high temperature slope of the mass plots is the same as for the LK predictions of $\beta+\alpha$. In other words, $m_{\beta-\alpha} \simeq m_{\alpha}+m_{\beta}$ at high temperature, in agreement with the numerical simulations of Ref.~\onlinecite{Fo05}. In contrast, strong deviations from the LK behaviour are observed at low temperature. Besides, as pointed out in  Ref.~\onlinecite{Gv04}, the amplitude $A_{\beta-\alpha}$ is higher than $A_{\beta+\alpha}$ in the case of  $\kappa$-(BEDT-TTF)$_2$Cu(NCS)$_2$. Again, this behaviour can be explained by the value of the spin damping factor which is smaller than for $\theta$-(BEDT-TTF)$_4$CoBr$_4$(C$_6$H$_4$Cl$_2$): $R^s_{\beta+\alpha}$ = -0.19 and -0.98, respectively.

\begin{figure}
\includegraphics[width=0.8\columnwidth,clip,angle=0]{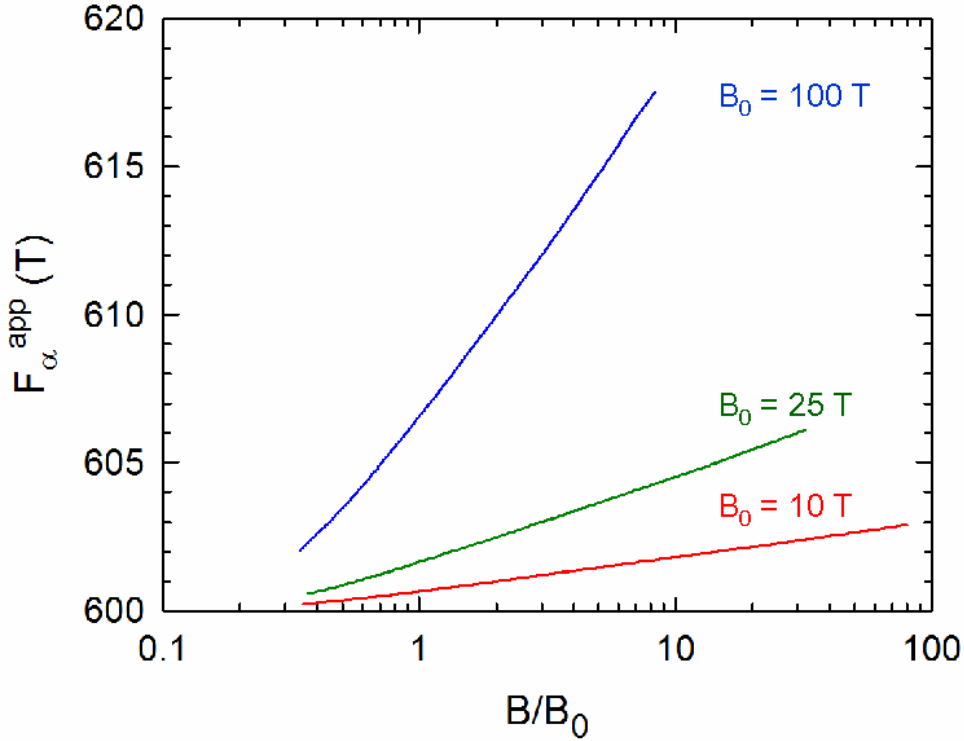}
\caption{\label{Fig:Fapp} Field dependence of the apparent frequency $F^{app}_{\alpha}$ deduced from Eq.~\ref{Eq:Fapp} for $F_{\alpha}$ = 600 T, relevant to  $\kappa$-(BEDT-TTF)$_2$Cu(NCS)$_2$, and various values of the magnetic breakdown field $B_0$.}
\end{figure}

Finally, let us consider briefly the field-dependent Onsager phase factor
introduced in Eq.~\ref{Eq:phi}. Since significant phase shift can be observed at
large magnetic field in Fig.~\ref{fig:omega}, the oscillation periodicity in 1/B
could be questioned. An 'apparent' oscillation frequency $F_{app}$ can be
defined as $1/F^{app}_{\eta}=1/B_i-1/B_{i+1}$ where $B_i$ and $B_{i+1}$ are the
fields at which two successive oscillation maxima occur. $F^{app}_{\eta}$ can be
evaluated through  an implicit equation deduced from Eq.~\ref{Eq:phi}:

\begin{equation}
\label{Eq:Fapp}
F^{app}_{\eta}=\frac{F_{\eta}}{1+n^r_{\eta}[\omega(B_i)-\omega(B_{i+1})]/2\pi}
\end{equation}

As reported in Fig.~\ref{Fig:Fapp} relevant to $F_{\alpha}$ for  $\kappa$-(BEDT-TTF)$_2$Cu(NCS)$_2$, frequency variations are small, even at very high field. As a result, owing to the limited field range in which oscillations are observed (even not to mention experimental uncertainties), the periodicity in $1/B$ is still observed and it can be checked that Fourier analysis yields clear peaks. These statements are in agreement with the data of Ref.~\onlinecite{Au13} relevant to $\theta$-(BEDT-TTF)$_4$CoBr$_4$(C$_6$H$_4$Cl$_2$).

\section{\label{sec:discussion}Summary and conclusion}

The field- and temperature-dependent amplitude and phase of de Haas-van Alphen oscillations relevant to the model Fermi surface by Pippard have been studied in the case of the organic metals  $\kappa$-(BEDT-TTF)$_2$Cu(NCS)$_2$ and $\theta$-(BEDT-TTF)$_4$CoBr$_4$(C$_6$H$_4$Cl$_2$). The main feature of the analytic formulaes governing the Fourier components amplitude is the presence of second order terms (the first order terms corresponding to the LK predictions).

Amplitude of the basic orbit  $\alpha$ and the MB-induced $\beta$ and
$2\beta-\alpha$ orbits, satisfactorily follow the LK behaviour, provided the
spin damping factor of the leading term is not too small, $i.e.$ far from
spin-zeroes. This result, due to small value of the second order terms,
validates the data analysis performed within the LK model, widely reported in
the literature. In contrast, besides deviations from the LK behaviour at low
temperature and high field, amplitude of the MB orbit  $\beta+\alpha$ and
harmonics may exhibit strong dips. Even though these dips are strongly linked to
the spin damping factors value, this behaviour is not due to spin-zero
phenomenon but to the cancellation of first and second order terms at peculiar
values of the magnetic field and temperature, instead. In particular, the
discrepancies observed in the reported data can be explained on the basis of
different orientation of the magnetic field with respect to the conducting plane
(different $\theta$ angle). At high temperature, the LK behaviour is observed for all the semiclassical orbits and harmonics. In particular, the effective mass follows the Falicov-Stachowiak model ($m_{n_{\alpha}\alpha+n_{\beta}\beta}=n_{\alpha}m_{\alpha}+n_{\beta}m_{\beta}$).

As for the 'forbidden orbit' $\beta-\alpha$, which is only governed by second order terms, although its Fourier amplitude strongly deviates from the LK behaviour at low temperature, its effective mass is given by $m_{\beta+\alpha}=m_{\alpha}+m_{\beta}$ at high $T/B$ ratio, in agreement with previous numerical simulations.

Finally, to have a full description of the oscillation spectrum, a field-dependent Onsager phase must be taken into account for Fourier components involving Bragg reflections. Nevertheless, although it can reach significant values, this additional phase have a little effect on the frequencies deduced from Fourier analysis.



\begin{thebibliography}{00}
\bibitem[17]{fn2} For convenience, energies ($E$) such as $\mu$, $\Delta_0$, $\Delta_1$ are expressed in units of
Tesla, using the conversion $(u_0/k_B)[T/J]\times  E[J]= E[T]$.
$B$ and $T$ are the magnetic field [T] and temperature [K], respectively.
Effective masses are expressed in units of the electron mass $m_e$, and magnetization
in Tesla units. The advantage of taking this convention is that field,
frequencies and temperature are not expressed in reduced units.
\bibitem[18]{fn1} In the case where the magnetic field direction is not parallel to the normal to the conducting plane (angle $\theta$), $B$ is changed to $B\cos\theta$ and the spin damping factor is written $R^{s}_{\eta,p} = \cos(\pi g_{\eta}m_{\eta}/2\cos\theta)$.
\bibitem[25]{fn3} Of course, dealing with real experiments, effective mass is merely deduced from fitting directly Eq.~\ref{Eq:y} to the mass plot data.
\bibitem[28]{fn4} Actually, parameters such as $F_{\eta}, B_0, T_{D\eta}$ are
pressure-dependent, as well, which influences the Fourier amplitude.
Nevertheless, the spin-zeroes angles only depend on the product
$g^*_{\eta}m_{\eta}$.
\end{thebibliography}

\end{document}